\documentclass[cernpreprint,texlive=2011,txfonts,UKenglish]{latex/atlasdoc}
\usepackage[backend=biber]{latex/atlaspackage}
\usepackage{latex/atlasbiblatex}
\usepackage{latex/atlasphysics}

\graphicspath{figures/}
\newcommand{\ddijetstitle}{{Dijet production in {\boldmath $\sqrt{s}=7$} \TeV\ 
{\boldmath $pp$} collisions with large rapidity gaps at the ATLAS experiment}}
\AtlasTitle{\ddijetstitle}

\author{The ATLAS Collaboration}

\PreprintIdNumber{CERN-PH-EP-2015-227}

\AtlasDate{\today}

 \AtlasJournal{Phys. Lett. B}
\AtlasAbstract{%
A $6.8 \ {\rm nb^{-1}}$ sample of $pp$ collision data collected under 
low-luminosity conditions
at $\sqrt{s} = 7$ \TeV\ by the ATLAS detector at the Large Hadron Collider is 
used to study diffractive dijet production. Events containing at least two jets 
with $p_\mathrm{T} > 20$ \GeV\ are selected and analysed in terms of 
variables which discriminate between diffractive and non-diffractive processes.
Cross sections are measured differentially in $\detaf$, the size of the 
observable forward region of pseudorapidity which is devoid of hadronic 
activity, and in an estimator, $\xiepz$, of the fractional momentum loss
of the proton assuming single diffractive dissociation ($pp \rightarrow pX$). 
Model comparisons indicate a
dominant non-diffractive contribution up to moderately large 
$\detaf$ and small $\xiepz$, with a diffractive contribution which 
is significant at the highest $\detaf$ and the lowest $\xiepz$. The 
rapidity-gap survival probability is estimated from comparisons of the data in 
this latter region with predictions based on diffractive parton distribution 
functions.}

\newcommand{\pom}{I$\!$P}

\newcommand{\lappr}{\stackrel{<}{_{\sim}}}

\newcommand{\xiepz}{\tilde\xi}

\newcommand{\detaf}{\ensuremath{\Delta\eta^{\mathrm{F}}}}

\newcommand{\degree}{\ensuremath{^\circ}}

\newcommand{\ptrans}{\ensuremath{p_\mathrm{T}}}

\newcommand{\lmodetal}[2]{\ensuremath{#1 < |\eta| < #2}}

\newcommand{\modetal}[1]{\ensuremath{|\eta| < #1}}

\newcommand{\antikt}{\ensuremath{\mathrm{anti}\mbox{-}k_{t}}}

\begin{document}
\nolinenumbers
\maketitle

%MT\tableofcontents

\section{Introduction}
Diffractive dissociation (e.g. $pp \rightarrow pX$) contributes a large 
fraction of the total inelastic cross section \cite{STDM-2010-11} at the Large 
Hadron Collider (LHC). The inclusive process has been studied using the 
earliest LHC data in samples of events in which a large gap is identified in 
the rapidity distribution of final-state hadrons 
\cite{STDM-2011-01,Khachatryan:2015gka}. In the absence of hard scales, the 
understanding of these data is based on phenomenological methods rather than 
the established theory of the strong interaction, quantum chromodynamics (QCD). 

A subset of diffractive dissociation events in which hadronic jets are produced
as components of the dissociation system, $X$, was first observed at the SPS 
\cite{Bonino:1988ae}, a phenomenon which has since been studied extensively at 
HERA \cite{Adloff:2000qi,Aaron:2010su} and the Tevatron \cite{Affolder:2000vb}. 
The jet transverse momentum provides a natural hard scale for perturbative QCD 
calculations, making the process sensitive to the underlying parton dynamics of 
diffraction and colour-singlet exchange. A model \cite{Ingelman:1984ns} in 
which the hard scattering is factorised from a colourless component of the 
proton with its own partonic content (diffractive parton distribution 
functions, DPDFs), corresponding to the older concept of a pomeron 
\cite{pomeron}, 
has been successful in describing diffractive deep inelastic scattering 
($ep \rightarrow eXp$) at HERA \cite{Newman:2013ada}. The DPDFs have been 
extracted from fits to HERA data in the framework of next-to-leading-order QCD, 
revealing a highly gluon-dominated structure 
\cite{Aktas:2006hy,Chekanov:2009aa}.

The success of the factorisable approach breaks down when DPDFs from 
$ep$ scattering are applied to hard diffractive cross sections in 
photoproduction \cite{Aktas:2007hn,Andreev:2015cwa} or at hadron colliders. 
Tevatron data \cite{Affolder:2000vb} 
show a suppression of the measured cross section 
by a factor of typically 10 relative to predictions.
A similar `rapidity-gap survival probability' factor, usually denoted by
$S^2$, was suggested by the first results from the 
LHC \cite{CMS-FWD-10-004}. This factorisation
breaking is usually attributed to secondary scattering from beam remnants,
also referred to as absorptive corrections, and closely related to the 
multiple-scattering effects which are a primary focus of 
underlying-event 
studies \cite{STDM-2010-04,STDM-2011-31,STDM-2011-42}. 
Understanding these effects more deeply
is an important step towards a complete model of diffractive processes
at hadronic colliders and may point the way towards a
reconciliation of the currently very
different theoretical treatments of soft and hard strong interactions. 

In this paper, the ATLAS technique for finding large rapidity gaps, first 
introduced in Ref. \cite{STDM-2011-01}, is developed further and applied
to events in which a pair of high transverse momentum ($\ptrans$) jets is 
identified.
The resulting cross sections are measured as a function of the size
of the rapidity gap and of an estimator
of the fractional energy loss of the intact proton. The results are
interpreted through comparisons with Monte Carlo models which incorporate
DPDF-based predictions with no modelling of multiple scattering. 
Comparisons between the measurements and the predictions thus provide
estimates of the rapidity-gap survival probability applicable to single
dissociation processes at LHC energies.

\section{Models and simulations}
\label{sec:MC}

Monte Carlo (MC) simulations using leading-order (LO) calculations in 
perturbative QCD are used in unfolding the data to correct for
experimental effects and in the comparison of the measurements with theoretical
models. The PYTHIA~8.165 (hereafter referred to as PYTHIA8) general-purpose 
LO MC generator \cite{Sjostrand:2007gs} is 
used to model dijet production in non-diffractive (ND) events, as well as in 
single diffractive dissociation (SD, $pp \rightarrow Xp$) and double 
diffractive dissociation (DD, $pp \rightarrow XY$). 
An alternative model of the SD process is provided by POMWIG (version 
2.0$\beta$) \cite{Cox:2000jt}, whilst an 
alternative next-to-leading-order (NLO) model of the ND process
is provided by POWHEG (version 1.0) \cite{Nason:2004rx,Alioli:2010xa}. 
\begin{figure}[h!]
  \begin{center}
\vspace*{0.5cm}
     \includegraphics[width=0.50\textwidth]{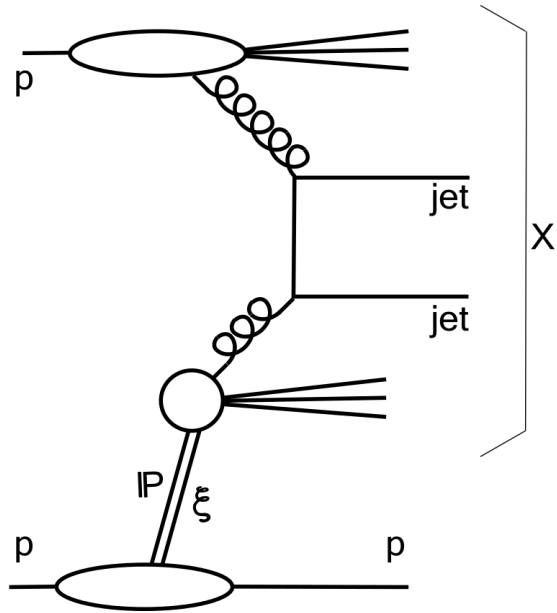}
\vspace*{-1.5cm}
    \caption{Illustration of hard single-diffractive scattering, in which 
partons from a pomeron (\pom) and from a proton enter a hard sub-process.
The rapidity gap appears between the system $X$ and the intact proton.}
    \label{fig:sdHardFeynman}
\vspace*{-0.5cm}
  \end{center}
\end{figure}

In both PYTHIA8 and POMWIG, hard scattering in diffractive processes takes 
place through the factorisable pomeron mechanism \cite{Ingelman:1984ns}
illustrated in Fig.~\ref{fig:sdHardFeynman}. 
A pomeron couples to an incoming proton, acquiring a fraction $\xi$ of the 
proton's longitudinal momentum. The proton either scatters elastically (SD) or
dissociates to form a higher-mass system (DD). A parton from the pomeron 
(as described by DPDFs) then undergoes a hard 
scattering with a parton from the dissociating proton at a scale set by the 
transverse momenta of the resulting jets. The dissociation system $X$ has an 
invariant mass $M_{\mathrm X}$, such that $\xi = M_{\mathrm X}^2 / s$ at 
a proton--proton centre-of-mass energy $\sqrt{s}$.

POMWIG is based on a standard implementation of hard diffractive scattering
with a factorisable pomeron, in which both the pomeron flux and the 
DPDFs are taken from the results of the H1 2006 DPDF fit B\footnote{The H1 Fit 
B DPDFs correspond to the sum of the SD process and the component of the DD 
process where the lower of the two proton dissociation masses is smaller than 
1.6 \GeV\ (see Section 6).} \cite{Aktas:2006hy}
and the proton PDF set is CTEQ61 \cite{Stump:2003yu}.
In contrast, PYTHIA8 provides a simultaneous model of hard and soft 
diffraction \cite{Navin:2010kk},
in which a soft diffractive model inherited 
from PYTHIA6 \cite{Sjostrand:2006za}
is smoothly interfaced to a hard diffractive model similar to that in POMWIG. 
The probability
of using the hard model depends on $M_X$.
The H1 2006 DPDF fit B is 
again used for
the partonic content of the pomeron and the proton 
partonic structure is taken from the CT10 
PDFs \cite{Lai:2010vv}. Several different 
pomeron flux parameterisations are 
available in PYTHIA8. 
In addition to the default Schuler and Sj\"ostrand (S-S) 
model \cite{Schuler:1993wr}, 
alternative parameterisations by Donnachie and 
Landshoff (D-L) \cite{Donnachie:1984xq}
and Berger and Streng \cite{Berger:1986iu,Streng:1988hd},
as well as the Minimum Bias Rockefeller (MBR) 
model \cite{Ciesielski:2012mc}, 
are also considered in this analysis. These models 
differ primarily in their predictions for the $\xi$ dependence 
of the cross section \cite{Navin:2010kk}. 
The DD process in PYTHIA8 is modelled similarly to the SD process. 
Neither of the diffractive models considered here 
take rapidity-gap destruction effects
into account, i.e. they set the rapidity gap survival probability 
$S^2 \equiv 1$.

An alternative for ND processes is provided by the POWHEG NLO 
generator. As described in Ref. \cite{Alioli:2010xa}, the `hardest emission cross 
section' approach used in POWHEG avoids the pathological behaviour observed 
in calculating cross sections with symmetric jet cuts in fixed-order NLO 
calculations. Here, NLO dijet production in the DGLAP formalism is interfaced with
PYTHIA 8 to resum soft and collinear emissions using the parton shower 
approximation.

PYTHIA8 adopts the Lund String model \cite{Andersson:1983ia} 
for hadronisation in each of the ND, SD and DD channels. 
It also contains an underlying-event model based on 
multiple parton interactions (MPI). 
POMWIG is derived from 
HERWIG \cite{Marchesini:1991ch} and
thus inherits its fragmentation and 
cluster-based hadronisation models.
For the purposes of this paper, the POWHEG ND simulation is
interfaced to 
PYTHIA8 for fragmentation and 
hadronisation.
All considered models based on the PYTHIA hadronisation model include 
$\ptrans$-ordered parton showering, while those based on HERWIG use 
angular-ordered parton showering.

The default MC combination used for the data unfolding for detector effects is 
a mixture of PYTHIA8 samples of ND, SD and DD dijets, with the 
``ATLAS AU2-CT10'' set of tuned parameters (tune) \cite{ATL-PHYS-PUB-2012-003}
for the underlying event.
In this tune, the fraction of the total cross section attributed to 
the SD process is reduced relative to the default by 10\% and that to DD 
by 12\%, to better match early LHC data. 
The Berger--Streng parameterisation, which has a very similar $\xi$
dependence to D-L, is chosen for the pomeron flux factor. Finally, the 
interaction of the particles with the ATLAS detector is simulated using a 
GEANT4-based program \cite{Agostinelli:2002hh,SOFT-2010-01}.

\section{The ATLAS detector}
\label{ch:atlasDetector}

The ATLAS detector is described in detail elsewhere \cite{PERF-2007-01}.
The beam-line is surrounded by a tracking
system, which covers the pseudorapidity\footnote{In the ATLAS coordinate system,
the $z$-axis points in the direction of the anti-clockwise beam viewed from 
above. Polar angles $\theta$ and transverse momenta \pt\ are measured with 
respect to this axis. 
The pseudorapidity $\eta = - \ln \, \tan( \theta / 2)$ 
is a good approximation to the rapidity of a particle 
whose mass is negligible compared with its energy and is used here, 
relative to the 
nominal $z = 0$ point at the centre of the apparatus, 
to describe regions of the detector.}
range \modetal{2.5}, consists of silicon pixel, silicon strip and straw
tube detectors and is immersed in the 2~T axial magnetic field of a 
superconducting solenoid. The calorimeters lie outside the tracking system. 
A highly 
segmented electromagnetic (EM) liquid-argon sampling calorimeter covers the 
range \modetal{3.2}. 
The EM calorimeter also includes a presampler
covering \modetal{1.8}. The hadronic end-cap 
(HEC, \lmodetal{1.5}{3.2}) and forward (FCAL, \lmodetal{3.1}{4.9}) 
calorimeters also use liquid argon for their sensitive layers, but 
with reduced granularity. Hadronic energy in the central
region is reconstructed in a steel/scintillator-tile 
calorimeter. The shapes of the cell noise distributions in
the calorimeters are well described by Gaussian distributions, 
with the exception of the tile calorimeter, where the noise 
has extended tails, and which is thus excluded from the 
rapidity gap finding aspects of the analysis.
Minimum-bias trigger scintillator (MBTS) detectors are 
mounted in front of the end-cap calorimeters on both sides of the 
interaction point and cover the pseudorapidity 
range \lmodetal{2.1}{3.8}. The MBTS is divided into 
inner and outer rings, both of which have eight-fold 
segmentation. 
In the analysis, two trigger systems are used at Level-1 (L1), namely the MBTS 
which efficiently collects low-$\ptrans$ jets, and the calorimeter-based trigger
(L1Calo) which concentrates on higher-$\ptrans$ jets. 
In 2010, the luminosity was measured by monitoring the activity in 
forward detector components, with 
calibration determined through van der 
Meer beam scans \cite{DAPR-2010-01, DAPR-2011-01}.

%%%%%%%%%%%%%%%%%%%%%%%%%%%%%%%%%%%%
%  ______               _      _____      _           _   _             
% |  ____|             | |    / ____|    | |         | | (_)            
% | |____   _____ _ __ | |_  | (___   ___| | ___  ___| |_ _  ___  _ __  
% |  __\ \ / / _ \ '_ \| __|  \___ \ / _ \ |/ _ \/ __| __| |/ _ \| '_ \ 
% | |___\ V /  __/ | | | |_   ____) |  __/ |  __/ (__| |_| | (_) | | | |
% |______\_/ \___|_| |_|\__| |_____/ \___|_|\___|\___|\__|_|\___/|_| |_|
%                                                                       
%%%%%%%%%%%%%%%%%%%%%%%%%%%%%%%%%%%%                                                                       

\section{Experimental method}
\label{ch:measurement}
To study rapidity-gap production, the experiment needs to operate at very low 
luminosities such that there is on average much less than one collision per 
bunch crossing (i.e. negligible 'pile-up'). This requirement has to be 
balanced against the need to collect adequate numbers of events with large 
rapidity gaps. The analysis therefore uses data from an early 2010 LHC run, 
with a total integrated luminosity of $6.8 \ {\rm nb^{-1}}$. The average number 
of collisions per bunch crossing is $0.12$. 

The jet selection follows that used in the ATLAS 2010 dijet analysis 
\cite{STDM-2011-02}. Jets with $p_{\mathrm T} > 20$ \GeV\ and $|\eta| < 4.4$
are reconstructed by applying the \antikt~algorithm \cite{Cacciari:2008gp} to 
topological clusters at the standard ATLAS jet energy scale. 
For comparisons, in particle-level MC models, 
jets are formed with the \antikt~algorithm from stable ($c\tau > 10$~mm)
final-state particles. The analysis is performed with jets of two different 
radius parameters $R = 0.4$ and $R = 0.6$. 
Approximately twice as many 
jets are reconstructed with the $R = 0.6$ than with the $R = 0.4$ requirement
in the kinematic range covered here. 

The calorimeter-based jet trigger ('L1Calo') is used
with the lowest available $p_{\mathrm T}$  
threshold in phase-space regions where
its efficiency is determined to be greater than 60\%. 
This criterion is satisfied for central jets at all pseudorapidities 
in the range \modetal{2.9} with $p_{\mathrm T} > 29\ (34)$ \GeV\ for 
jets with $R = 0.4\ (0.6)$. At lower transverse momenta, or where
the jets are beyond the L1Calo $\eta$ range, 
the MBTS trigger is used, with the requirement of a signal in
at least one segment. The MBTS trigger is fully efficient for dijet 
events, but
has a substantial time-dependent
prescale (which is taken into account in the 
off-line analysis), reducing the effective luminosity for forward and 
low-$p_{\mathrm T}$ jets to $0.303 \ {\rm nb^{-1}}$.

At least two jets are required, with jet barycentres 
satisfying \modetal{4.4} and with $p_{\mathrm T} > 20$ \GeV.
These requirements correspond to the region in 
which the jet energy scale and resolution are well known and in which the
jets are fully contained within the detector. 

Several sources of background were investigated. To reject contributions 
from beam interactions with residual gas in the beampipe, muons from
upstream proton interactions travelling as a halo around the proton beam, and 
cosmic-ray muons, events are required to have a primary vertex constructed from 
at least two tracks and consistent with the beam spot position. 
In-time pile-up, caused by multiple interactions in one bunch crossing,
is suppressed by requiring that there be no further vertices with two or more 
associated tracks. 
Out-of-time pile-up, caused by overlapping signals in the detector from 
neighbouring bunch crossings, was investigated and  
found to be negligible at the large bunch spacings 
($> 5 \ {\rm \mu s}$) of the chosen runs. 
Once an event is triggered and the dijet selection criteria
are met, the requirement on the primary vertex removes 0.3\% and
0.2\% of events in the L1Calo- and MBTS-triggered data, respectively, while
the in-time
pile-up suppression cuts remove 9.4\% and 6.5\%, respectively. The latter 
values are used to scale the cross sections to account for 
the corresponding losses.
Residual background occurs due to the limited position resolution
of the vertex reconstruction,
which typically merges pairs of vertices with 
$\Delta z \lesssim 1 \ {\rm cm}$ 
into a single vertex. The size of this effect
is estimated by extrapolation 
to lower values of the
$\Delta z$ distribution for pairs of vertices which are resolved
and its influence is evaluated
by randomly overlaying minimum-bias events on the selected sample.  
The effect is smaller than 0.5\% in all bins of the measured distributions. 
The residual beam-induced background is studied using 
`unpaired' bunch crossings in which only 
one bunch of protons passes through the ATLAS detector and is found to be 
negligible.

Each event is characterised in terms of pseudorapidity regions which are devoid 
of hadronic activity ('rapidity gaps') using a method very similar to that 
first introduced in Ref. \cite{STDM-2011-01}. Rapidity gaps are defined using 
the tracking (\modetal{2.5} and $p_{\mathrm T} > 200$ \MeV) and calorimetric 
(\modetal{4.8}) information within the ATLAS detector acceptance.
Full details of the track selection can be found in Ref. \cite{STDM-2010-06}. 
Following Ref. \cite{STDM-2011-01}, the clustering algorithm accepts 
calorimeter cells as cluster seeds if their measured response is approximately
five standard deviations above the root-mean-square noise level, with a small 
dependence of the threshold on pseudorapidity. Cells neighbouring the seed cell
are included in the cluster if their measured energies exceed smaller threshold
requirements defined by the standard ATLAS topological clustering method.
The particle-level gap definition is determined by the region of pseudorapidity
with an absence of neutral particles with $p > 200$ \MeV\ and charged particles
with either $p > 500$ \MeV\ or $\ptrans > 200$ \MeV. These momentum and 
transverse
momentum requirements match the ranges over which the simulation indicates that
particles are likely to be recorded in the detectors, accounting for the axial
magnetic field in the inner detector. The treatment of 
calorimeter information in the rapidity-gap determination follows the procedure 
introduced in Ref. \cite{STDM-2011-44}, such that the requirement 
$\ptrans > 200$ \MeV\ for calorimeter clusters from the previous rapidity-gap 
analysis \cite{STDM-2011-01} is removed. Since this transverse momentum 
requirement corresponds to a very high momentum at large pseudorapidities, the 
modified approach more completely exploits the capabilities of ATLAS to detect 
low-momentum particles in the calorimeters. The total numbers of selected events
in the L1Calo and MBTS samples with $R = 0.6$ are 285191 and 44372, 
respectively.

The variable characterising forward rapidity gaps, $\detaf$, is defined by the 
larger of the two empty pseudorapidity regions extending between the edges 
of the detector acceptance at $\eta=4.8$ or $\eta=-4.8$
and the nearest track or calorimeter cluster passing the selection
requirements at smaller $| \eta |$.
No requirements are placed on particle production at $|\eta| > 4.8$
and no attempt is made to identify gaps in the central
region of the detector. In this analysis,
the size of the rapidity gap relative to $\eta = \pm 4.8$
lies in the range $0 < \detaf < 6.5$. For example
$\detaf = 6.5$ implies that there is no 
reconstructed particle with (transverse) momentum  
above threshold in one of the regions $-4.8 < \eta < 1.7$ or
$-1.7 < \eta < 4.8$. 

For events which are of diffractive origin, the Monte Carlo studies 
indicate that the rapidity-gap definition selects processes in which one of the 
incoming protons either remains intact (SD) or is excited to produce a system 
with mass 
$M < 7$ \GeV\ (DD).
In the second case, the system is typically restricted to a pseudorapidity 
region beyond the acceptance 
of the ATLAS detector. 
In both cases, the other incoming proton dissociates to produce a 
hadronic system of larger invariant mass $M_{\mathrm X}$. The gap size, $\detaf$, 
grows approximately logarithmically with $1/M_{\mathrm X}$, the degree of 
correlation being limited by event-to-event hadronisation fluctuations. 

\begin{figure}[h!]
\begin{center}
\includegraphics[width=0.48\textwidth]{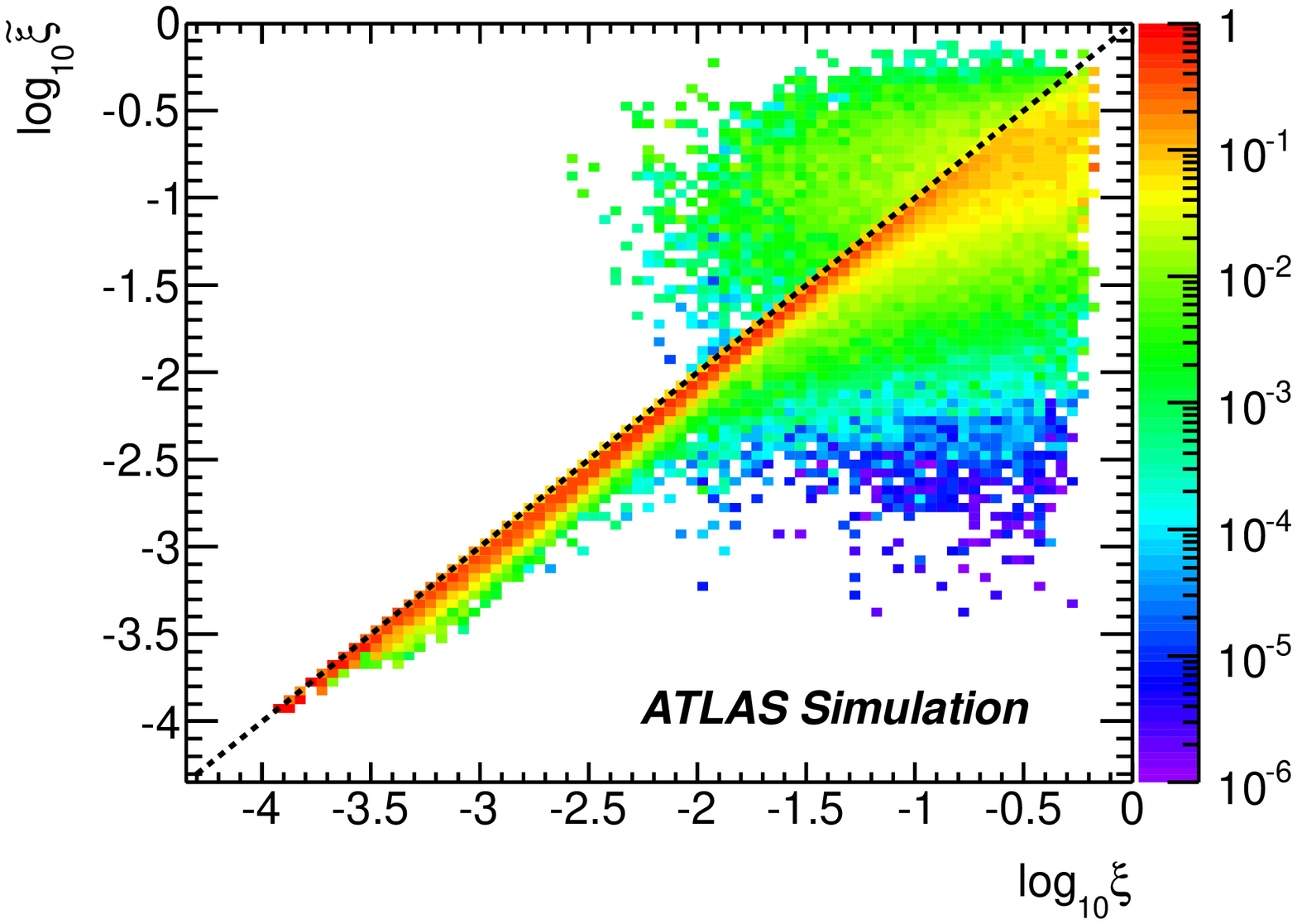}  
\includegraphics[width=0.48\textwidth]{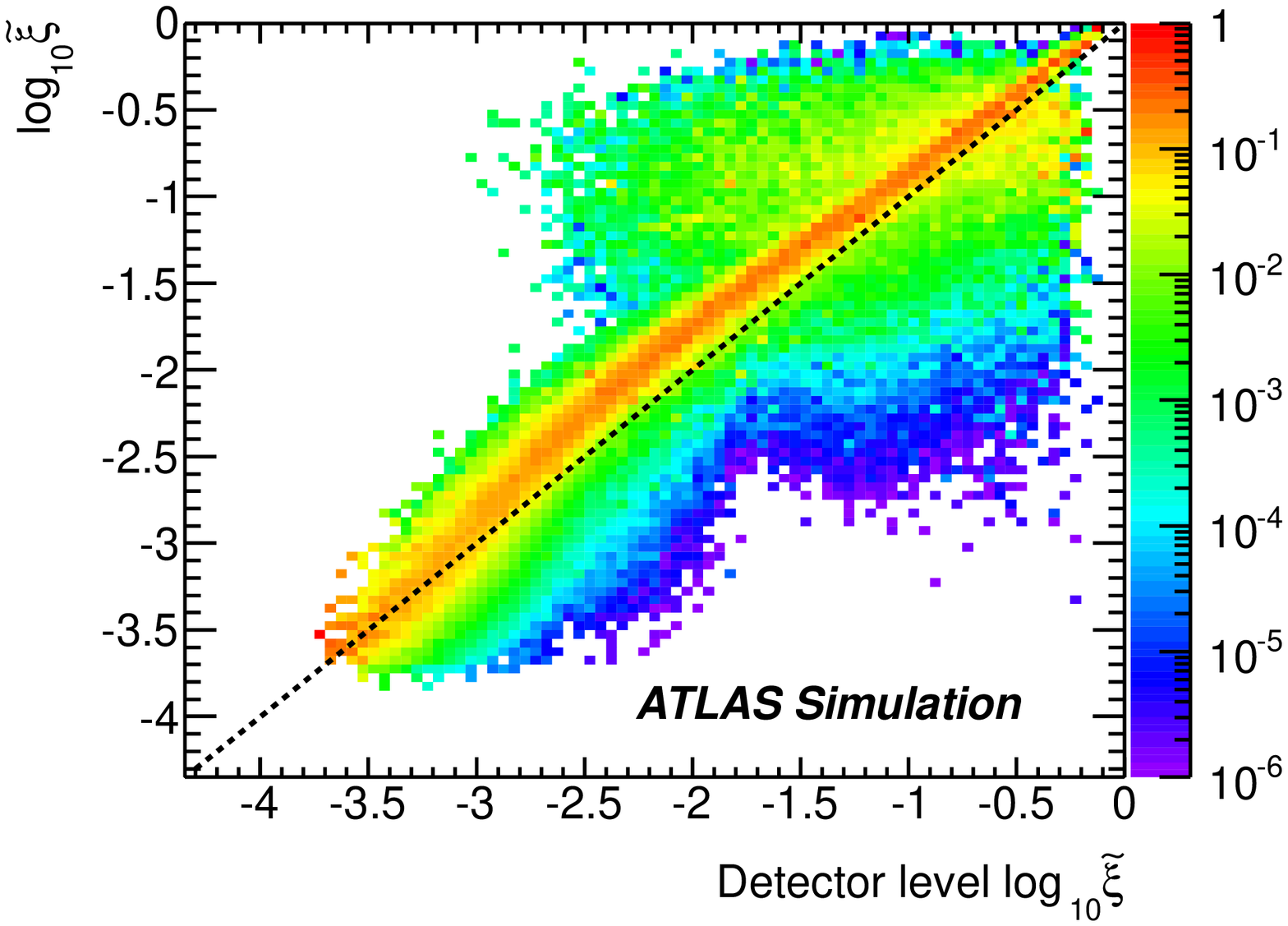}
  \caption{(a) Particle-level correlation between the $\xi$ variable extracted
from the diffractively scattered proton and $\xiepz$ calculated from particles
selected as defined in the text, using the PYTHIA8 SD MC model. (b) Correlation
between the particle-level $\xiepz$ and detector-level $\xiepz$ calculated
from clusters selected as defined in the text, using the sum of 
PYTHIA8 ND, SD and DD contributions. In both plots, the distributions
are normalised to unity in each column.}
  \label{fig:xicorrelations}
\end{center}
\end{figure}

In this analysis, measurements of the energy deposits in each event are used to 
construct a variable, $\xiepz$ which is closely correlated with $\xi$ and is 
similar to that used in Ref. \cite{CMS-FWD-10-004}. Neglecting any overall 
transverse momentum of the system $X$, the relation 
\begin{equation}
 \label{eq:approximateMxSq}
 M_{\mathrm X}^2 = \sqrt{s} \sum\ptrans e^{\pm\eta} \ ,
\end{equation}
holds for cases where the intact proton travels in the $\pm z$ direction.
In other words, if the forward rapidity gap starts at $\eta = +4.8$ (-4.8), 
the exponential function takes the positive (negative) sign. 
Here, the sum runs over all particles constituting the system $X$. 
This relation has the attractive feature that 
the sum is relatively insensitive to particles in the $X$ system 
travelling in the very forward direction, i.e. those which are
produced at large pseudorapidities beyond the detector acceptance.
Correspondingly, the variable $\xiepz$ is defined as 
\begin{equation}
 \label{eq:fiducialXi}
 \xiepz \simeq M_{\mathrm X}^2/s = \sum\ptrans e^{\pm\eta} / \sqrt{s} \ .
\end{equation} 
At the detector level, the sum in Eq.(\ref{eq:fiducialXi}) runs over 
calorimeter clusters in the region \modetal{4.8}. To best match this
requirement, the corrected cross section is defined
in terms of neutral particles with $p > 200$ \MeV\ and charged particles with 
$p > 500$ \MeV\ in the same pseudorapidity 
range. The correlation at the particle level between $\xiepz$ and the true 
$\xi$ (the latter obtained from elastically scattered protons) in the PYTHIA8 
MC model of SD events with two jets, is shown in 
Fig.~\ref{fig:xicorrelations}(a). 
For $\log_{10}\xi \lappr -2$, there is a clear correlation between the 
fiducial $\xiepz$ variable and $\xi$, which continues to larger $\xi$,
but with a progressively worse correspondence as some components of the
dissociation system which are included in the $\xi$ calculation fail the
fiducial requirement $|\eta| < 4.8$ applied in the $\xiepz$ calculation. 
At low values, $\xiepz$ is systematically slightly smaller than $\xi$, 
due to the exclusion of low-momentum particles from the $\xiepz$ definition.
Figure \ref{fig:xicorrelations}(b) shows the correlation between 
the reconstructed and particle-level determinations of 
$\xiepz$.
According to the MC models, the resolution in the absolute value of
$\log_{10} \xiepz$ varies from around 0.07 at large $\xiepz$ values to around 
0.14 at small $\xiepz$.

\begin{figure}[h!]
  \begin{center}
        \includegraphics[width=0.49\textwidth]{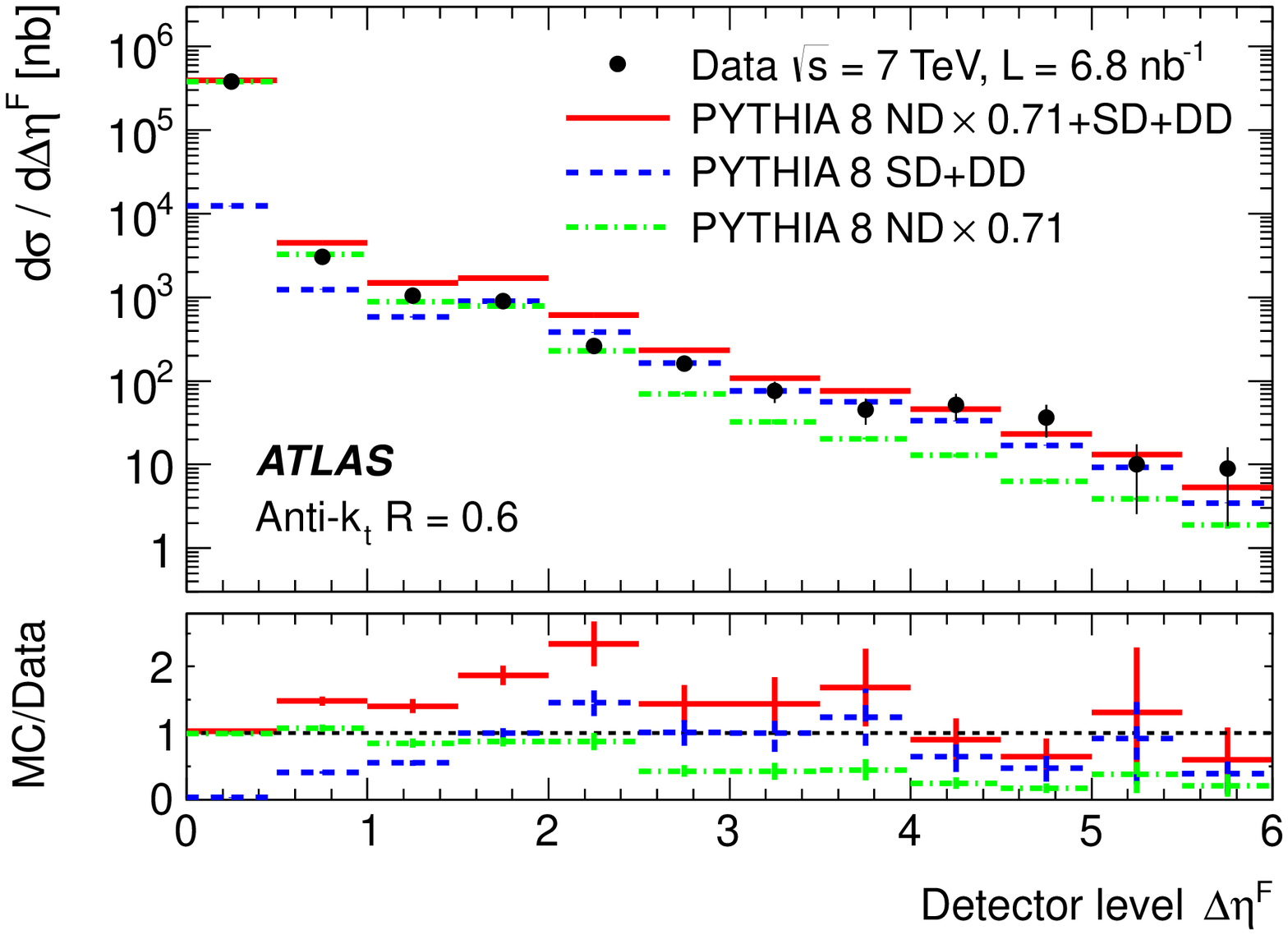} 
        \includegraphics[width=0.49\textwidth]{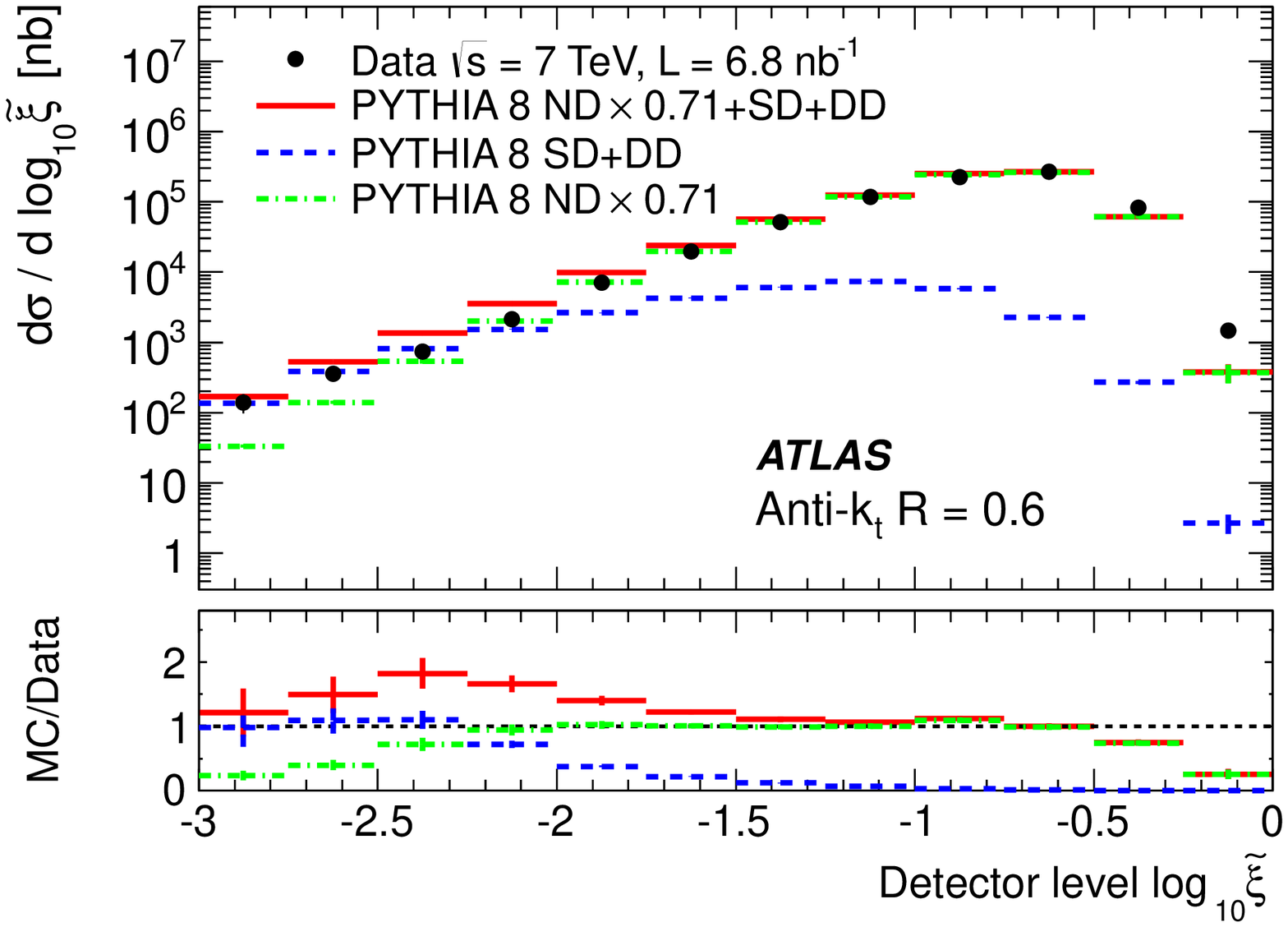}
        \includegraphics[width=0.49\textwidth]{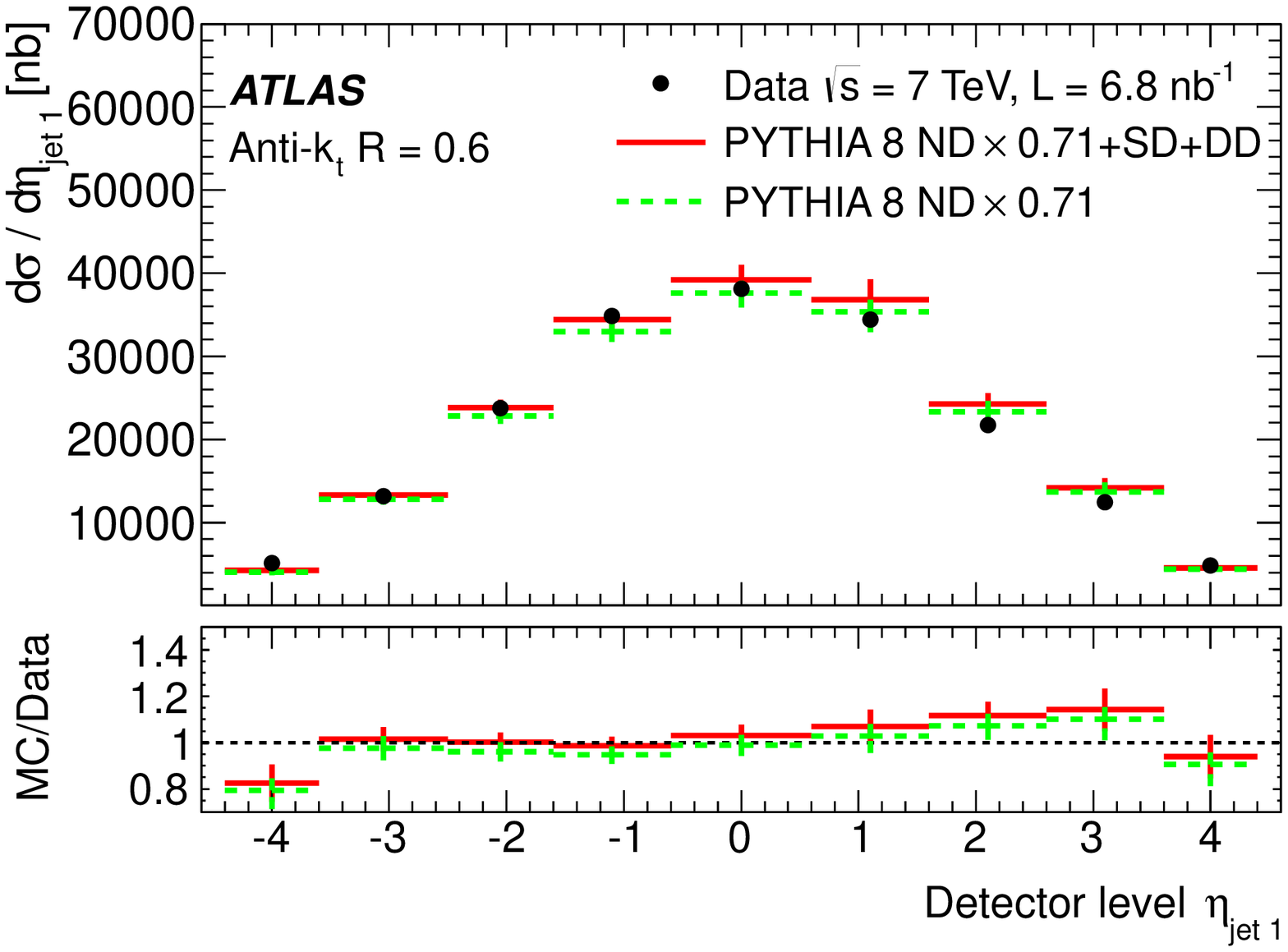}
        \includegraphics[width=0.49\textwidth]{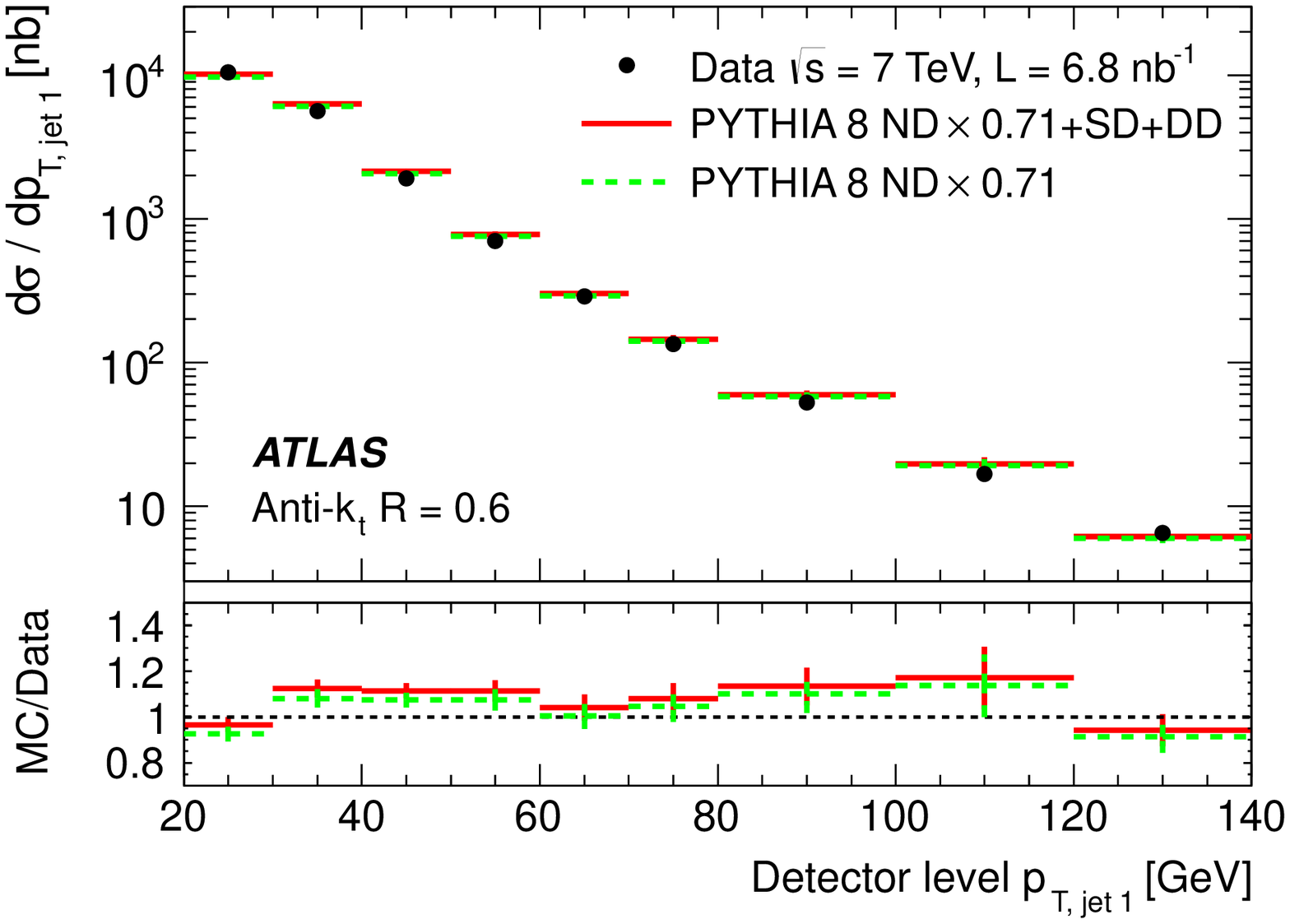}
  \caption{ 
Comparisons of dijet cross sections from uncorrected data with a combination of 
PYTHIA8 diffractive and non-diffractive contributions at detector level based 
on jets found by the \antikt\ algorithm with $R=0.6$. 
The MC distributions are normalised to the integrated luminosity of the data 
after first applying a factor of $0.71$ to the ND contribution. The error bars 
correspond to the statistical uncertainties. In addition to the measured 
(a) $\detaf$ and (b) $\xiepz$ variables, the distributions in (c) the 
leading-jet pseudorapidity and (d) transverse momentum are also shown.
The lower panels show ratios of the MC models to the data where the error bars 
indicate the sum in quadrature of the statistical uncertainties arising from 
the data and the MC simulation.}
 \label{fig:controlplots}
\end{center}
\end{figure}

The quality of the description of the uncorrected data by 
the PYTHIA8 Monte Carlo model is shown for several 
variables in Fig.~\ref{fig:controlplots}. Here, the default
ND component of PYTHIA8 is fixed to match the data in the first bin of
the $\detaf$ distribution, requiring a normalisation factor of $0.71$. 
The SD and DD contributions are shown without any adjustment of their
normalisation. Satisfactory descriptions are obtained of the $\detaf$ and 
$\xiepz$ variables, and also of the 
pseudorapidity and transverse momentum distributions of the leading 
jet, indicating that a combination of the diffractive and the non-diffractive
PYTHIA8 components is appropriate for use in the unfolding of
experimental effects. 

The data distributions in $\detaf$ and $\xiepz$ are corrected for detector 
acceptance and migrations between measurement bins due to finite experimental 
resolution using Iterative Dynamically Stabilised (IDS) unfolding 
\cite{Malaescu:2011yg}. 
This procedure corrects for migrations between the particle and detector
levels based on an `unfolding' matrix, constructed from a combination of 
PYTHIA8 ND, SD and DD samples, as shown in Fig.~\ref{fig:xicorrelations}(b).
The MC combination is optimised in a simple fitting procedure in which 
scaling factors are applied to the ND and (SD+DD) components to best match
the data. The IDS unfolding is performed in two dimensions, corresponding to 
the $\ptrans$ of the leading jet and the target distribution (either $\detaf$ 
or $\xiepz$). The results of the IDS procedure depend in general on the number 
of iterations used. A fast convergence is achieved for both measured 
distributions and the fourth iteration is chosen as nominal since it optimises 
the balance between the systematic and statistical uncertainty arising from the 
unfolding procedure. 
The unfolding procedure is stable against variations in 
binning, number of iterations and the scaling factors applied to 
the diffractive and non-diffractive contributions in the PYTHIA8 model,
as discussed further in Section~\ref{ses:Syst}.

\section{Systematic uncertainties}  
\label{ses:Syst}
The procedures for handling
many of the sources of systematic uncertainty  
follow from previous ATLAS measurements. 
The full list of uncertainties 
considered is given below.
Further details of the 
uncertainties affecting jets (sources 1--5 below)
can be found in  
Ref.~\cite{STDM-2011-02},
while those affecting diffractive variables (sources 7--9) are elaborated in 
Ref.~\cite{STDM-2011-01,STDM-2011-44}. 

\begin{enumerate}
\item {\bf Jet energy scale:} the largest source of uncertainty arises from the 
determination of the jet energy scale. This is obtained 
following the procedure in Ref. \cite{STDM-2011-02}, where relative shifts
are applied between the particle-level and detector-level response 
as a 
function of $\eta$ and $\ptrans$. 
This accounts for all
effects playing a role in evaluating jet transverse momenta,
including
dead material, 
electronic noise, the different responses of the LAr and 
Tile calorimeters, the 
simulation of particle showers in the calorimeters, pile-up effects and the 
models of fragmentation used by different MC generators 
\cite{PERF-2012-01}.
Studies in the context of the current analysis show that the
inclusive treatment is also appropriate for diffractive processes. 
As in Ref.
\cite{STDM-2011-02}, the dominant component of this uncertainty comes from the 
inter-calibration of jets in $\eta$.
The total resulting uncertainty in 
the differential cross sections measured here
varies from $20\%$ for small gaps to 
$\sim 40\%$ for very large gaps, a region which is dominated by
diffractive events with relatively
small transverse momentum or large pseudorapidity of jets. 
\item {\bf Jet energy resolution:} this is determined from data 
using in situ techniques and MC simulation \cite{PERF-2011-04}. The resulting 
uncertainty on the cross-section measurements is evaluated by smearing the 
$\ptrans$ of the reconstructed jets in MC simulation using a Gaussian 
distribution to match
the resolution uncertainty found in data. The resulting effect is 
below 6\% in all kinematic regions.
\item {\bf Jet angular resolution:} this was determined using
the same techniques as for the jet energy resolution. 
Following the procedure in Ref. 
\cite{STDM-2011-02}
leads to an uncertainty on the 
differential cross sections which is typically around 1--2\% and
largest for 
jets at the largest $|\eta|$.
\item {\bf Jet reconstruction efficiency:} the 
efficiency for reconstructing jets 
from the calorimeter information is determined by reference to a sample 
of `track jets' reconstructed from inner-detector tracks.
Following Ref.~\cite{STDM-2011-02}, the uncertainty is taken from the 
difference between the results of this procedure
using data and MC simulation, with extrapolation to the $\eta$ range
not covered by the tracker. This
results in systematic uncertainties in the measured
cross sections which are smaller than 2\% in all kinematic regions.
\item {\bf Jet cleaning efficiency:} the fraction of jets that match the 
standard quality criteria, designed to remove 
jets associated with spurious calorimeter response, 
was studied using a tag-and-probe technique \cite{STDM-2011-02}.  
The corresponding
systematic uncertainties 
are obtained by 
applying looser and tighter selections to the tag jet
and propagate to at most 8\% in the cross sections measured here. 
\item {\bf Trigger efficiency:} the trigger efficiency is evaluated as a 
function of leading-jet transverse momentum in various pseudorapidity ranges 
using either an independently triggered data sample or the MC mixture used in 
Fig.~\ref{fig:controlplots}. The rise near the threshold of the efficiency in each 
$\ptrans$ interval is parameterised based on a fit with free 
parameters. The efficiency is taken from the data, while the uncertainty is
taken as the difference between two MC distributions: one assuming 100\% trigger
efficiency and the other rescaled by trigger efficiencies found in this MC 
sample in the same $\eta$ and $\ptrans$ ranges as in the data. The resulting 
uncertainties are smaller  
than 3.5\% for 
all measured bins. A further parameterisation uncertainty, evaluated by varying 
the fit parameters within their uncertainties, is less than 0.7\% for all 
measurements. 
An additional uncertainty, below 0.5\% in all bins, is obtained from the 
differences in the simulated efficiencies from the ND, SD and DD processes.
\item {\bf Cluster energy scale:} the uncertainty on the energy scale of the 
individual calorimeter clusters used to determine $\xiepz$ is 
evaluated
in an $\eta$-dependent manner as described in Ref. \cite{STDM-2011-44}. 
The resulting uncertainty in the cross sections differential in $\xiepz$
is typically 10\%. 
\item {\bf Cell significance threshold:} 
the significance thresholds applied
to suppress calorimeter clusters 
which are consistent with noise fluctuations, 
are shifted up and down by 10\% to determine the corresponding
systematic uncertainties.
The weakened requirements on particle (transverse) momenta 
applied here compared with Ref.~\cite{STDM-2011-01} increase the
sensitivity to the threshold shifts, 
particularly in the forward regions, resulting in uncertainties on the 
differential cross sections of typically 10--20\%.
\item {\bf Track reconstruction efficiency:} the uncertainty on the track 
reconstruction efficiency is taken from Ref.~\cite{STDM-2010-06},
resulting in a negligible effect on the differential cross sections.
\item {\bf Luminosity:} the uncertainty on the luminosity is taken from the 
luminosity determination for the year 2010 \cite{DAPR-2011-01}, resulting in a 
$\pm$3.5\%
normalisation uncertainty on all measurements.
\item {\bf Reconstructed vertex requirement:} 
the uncertainty on the efficiency of the vertex multiplicity requirement is 
evaluated by loosening it in data to include events with no 
vertices. This 
changes the differential cross sections by less than 1\% in all bins.
\item {\bf Dead material:} the effect of possible inaccuracies in the detector 
dead material simulation
was studied in Ref.~\cite{STDM-2011-01} using dedicated MC samples with 
modified 
material budgets ($\pm$10\% around the central value) in
the inner detector, services and calorimeters. The largest effect 
on any bin in that analysis 
was 3\%, which is
applied as a symmetric shift in each bin of the 
current measurement. 
\item {\bf Unfolding procedure:} the uncertainty associated with modelling bias
introduced by the unfolding procedure is estimated using a data-driven procedure
whereby 
the particle-level distributions of the MC sample are reweighted such that the 
corresponding detector-level distributions match the uncorrected 
data in the two-dimensional ($\detaf$, $\xiepz$)-space. The reweighted 
detector-level MC distribution is then unfolded using the same procedure as is 
applied to the data. The systematic uncertainty in each bin is taken to be 
the  
difference between the unfolded reweighted MC distribution and the reweighted 
particle-level MC distribution. 
The resulting unfolding uncertainty is typically around 15\% for the 
$\detaf$ distribution (rising to 25\% in the bin for the largest gaps) and 
is smaller than 10\% in the case of the $\xiepz$ distribution. 
Since the factors used to scale the ND and (SD+DD) processes to best describe
the data before unfolding are different for the $\detaf$ and $\xiepz$ 
distributions, a further uncertainty of up to around 5\% is ascribed by swapping
these factors between the two distributions. 
\end{enumerate}

The total systematic uncertainty is defined as the sum in quadrature 
of the uncertainties described above.
The dominant contribution 
arises from the jet energy scale uncertainty, 
followed by the unfolding uncertainty, the 
cell significance threshold uncertainty (for the $\detaf$ distribution) 
and the cluster 
energy scale uncertainty (for $\xiepz$). The overall 
uncertainty varies between bins in the range 20\% to 45\%.
There are strong correlations between the systematic uncertainties in 
neighbouring measurement intervals of both the $\detaf$ and $\xiepz$ 
distributions. 

\section{Results}
\label{ch:results}

In this section, particle-level
dijet cross sections are presented differentially
in the variables \detaf and $\xiepz$, both of which have discriminatory
power to separate diffractive and non-diffractive contributions.
The cross sections correspond to events with at least two jets
with $\ptrans > 20$ \GeV\ in the region \modetal{4.4}. 
The particle-level gap is defined by the region of pseudorapidity
with an absence of neutral particles with $p > 200$~MeV and charged particles
with either $p > 500$ \MeV\ or $\ptrans > 200$ \MeV.
The conclusions are not strongly
dependent on the choice of $R$ parameter in the \antikt\ jet algorithm,
although the cross-section normalisations are about two times larger for 
$R = 0.6$ than for $R = 0.4$. 
The data shown here correspond to $R = 0.6$. The results with both
cone sizes can be found 
in tabular form in Ref. \cite{hepdata}.

\begin{figure}[h!]
  \begin{center}
\includegraphics[width=0.49\textwidth]{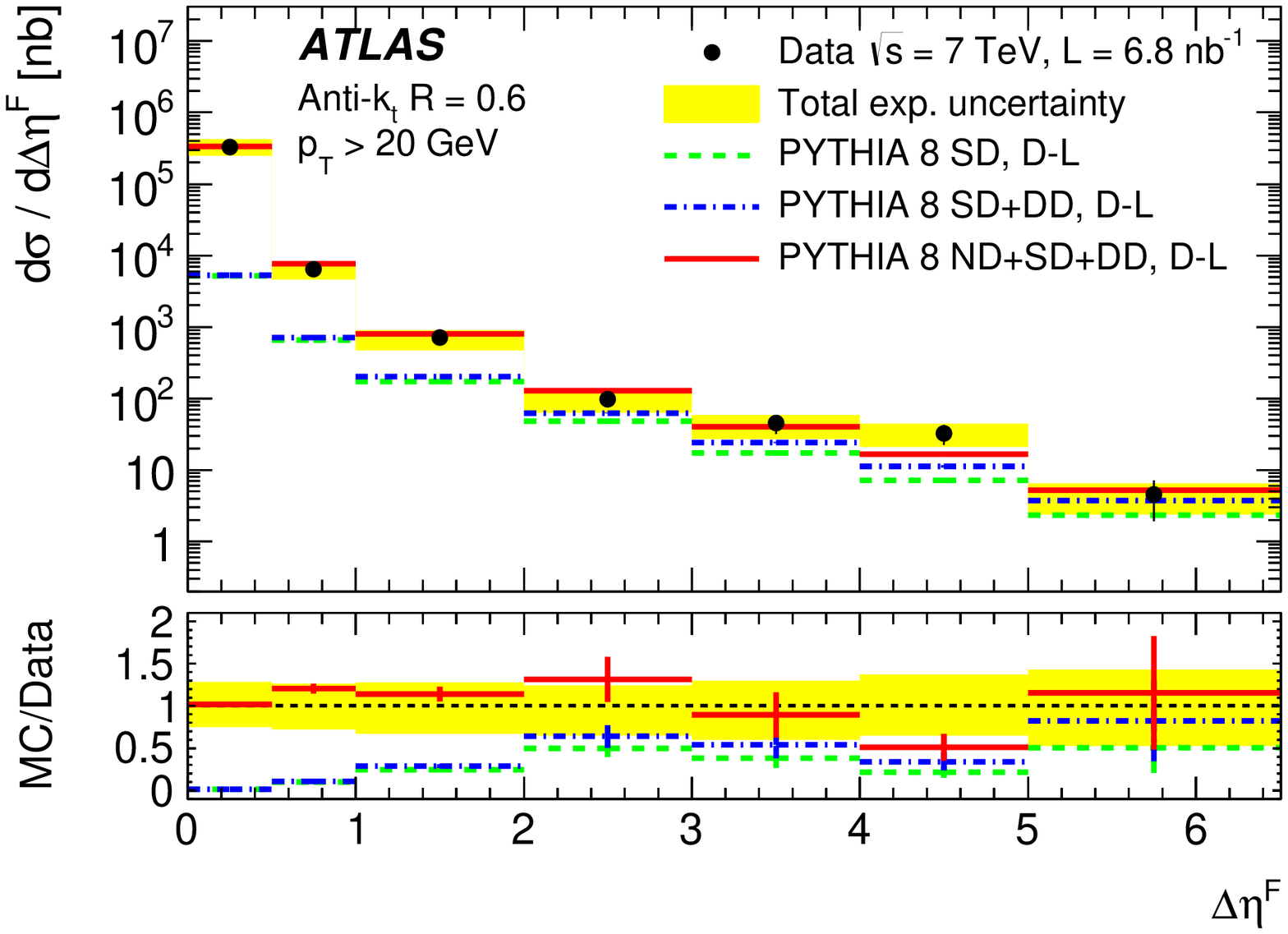}  
\includegraphics[width=0.49\textwidth]{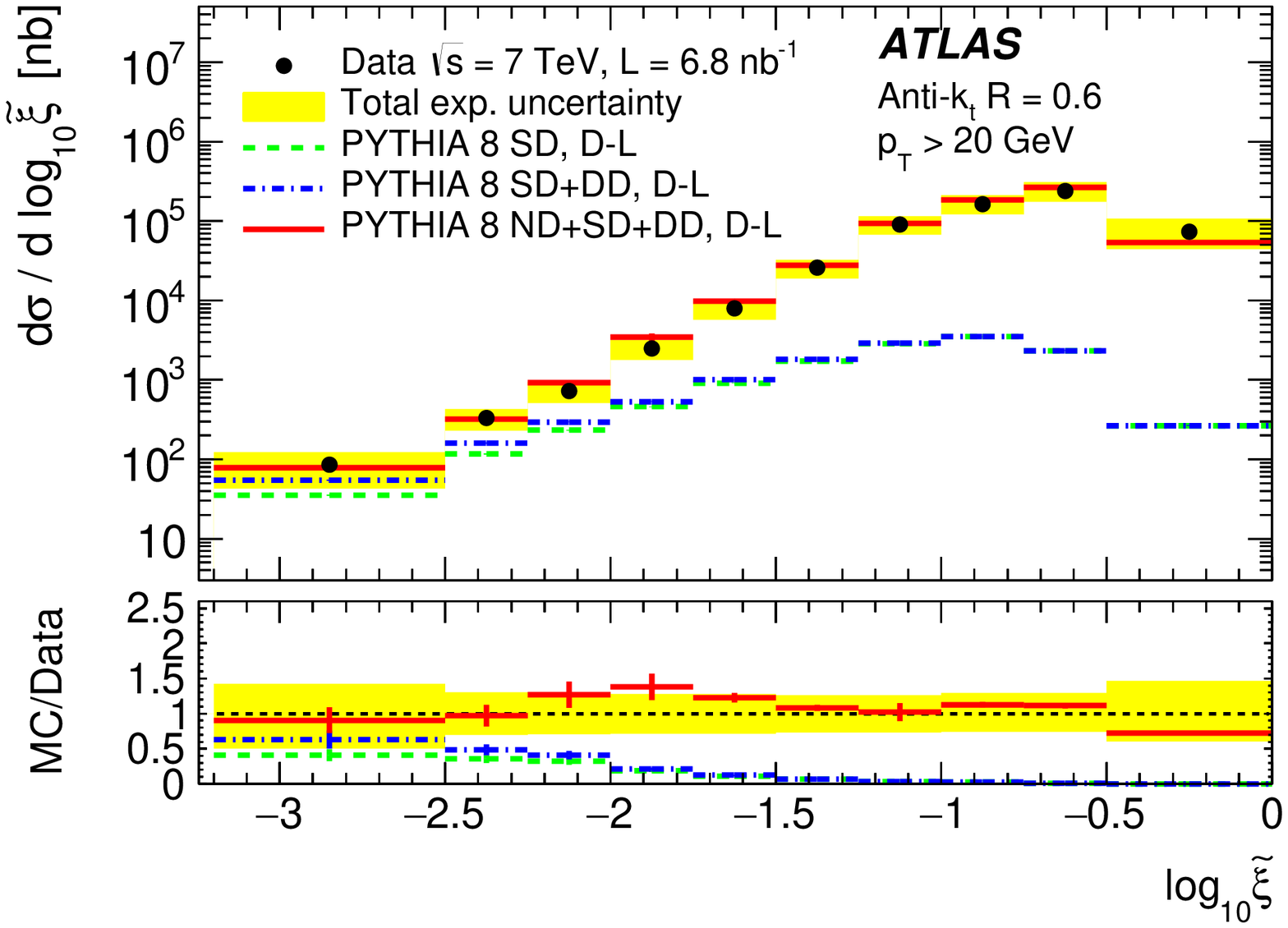}
    \caption{The differential dijet cross sections in (a) $\detaf$ and (b) 
$\xiepz$, compared with the particle-level PYTHIA8 model of the SD, sum of 
diffractive components SD and DD, and sum of all three ND, SD and DD components.
The Donnachie--Landshoff pomeron flux model is used for the diffractive 
components. The error bars on the data and the MC models indicate their 
respective statistical uncertainties, while the yellow bands show the total 
uncertainties on the data. The ND contribution is normalised to match the data 
in the first $\detaf$ bin. The lower panels show ratios of the MC models to the 
data where the error bars indicate the sum in quadrature of the statistical 
uncertainties arising from the data and the MC simulation.}
  \label{fig:results_Total_componentsStack}
\end{center}
\end{figure}

Figures \ref{fig:results_Total_componentsStack}(a) and
\ref{fig:results_Total_componentsStack}(b) show the dijet cross section 
differentially in \detaf and $\xiepz$ for $R = 0.6$ jets.
In contrast to
related distributions in inclusive rapidity-gap 
measurements \cite{STDM-2011-01}, 
the data in these figures do not 
show any significant diffractive plateau at large gap sizes. 
This difference is of kinematic origin, resulting from the reduced phase
space at large gap sizes or small $\xiepz$ when high-$p_{\mathrm T}$ 
jets are required. 
Both distributions are compared with predictions from the PYTHIA8
MC model, decomposed into ND, SD and DD components, with the D-L flux choice.
The normalisation of the 
ND contribution in both distributions is fixed to match the data in the 
first bin of \detaf, where this component is expected to 
be heavily dominant, requiring a multiplicative factor of 1/1.4. 
The SD and DD normalisations are left unchanged from their
defaults in PYTHIA8. This MC combination results in a satisfactory description 
of both distributions.
The ND component is at least an order of magnitude larger than 
the SD and DD contributions for relatively small 
$\detaf\lesssim 1$ and large 
$\xiepz \gtrsim 0.1$. As $\detaf$ grows or $\xiepz$ falls, the diffractive
components of the models become increasingly important,  
such that the ND and (SD+DD) components are  approximately equal at
$\detaf\sim 3$ or $\log_{10}\xiepz \sim -2$. At the largest gaps 
($\detaf \gtrsim 5$) and smallest $\xiepz$ ($\xiepz \lesssim 0.003$), 
the model suggests that the diffractive components are approximately 
twice as large as the ND contribution. 

A dijet cross section differential in $\xiepz$ has also 
been measured by CMS 
\cite{CMS-FWD-10-004}. 
The ATLAS and CMS hadron level cross-section
definitions are slightly different in terms of the 
$\eta$, $p$ and $\ptrans$ ranges of the particles considered
and the jet $R$ parameter.
Nonetheless, the measured cross sections are similar in magnitude
and both analyses lead to the conclusion that 
a non-negligible ND contribution 
extends to relatively large $\detaf$ and small $\xiepz$.

\begin{figure}[h!]
  \begin{center}
\includegraphics[width=0.49\textwidth]{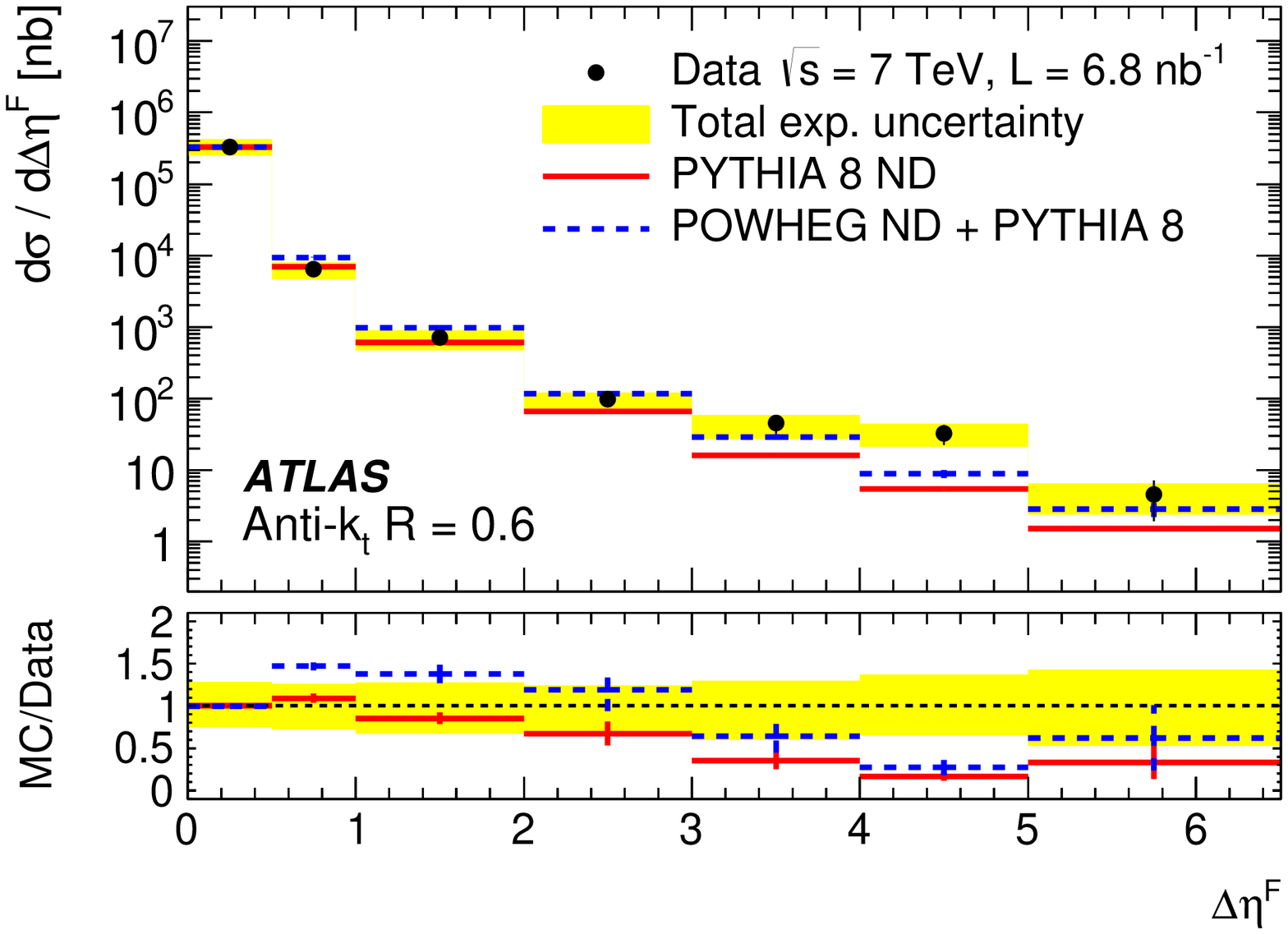}           
\includegraphics[width=0.49\textwidth]{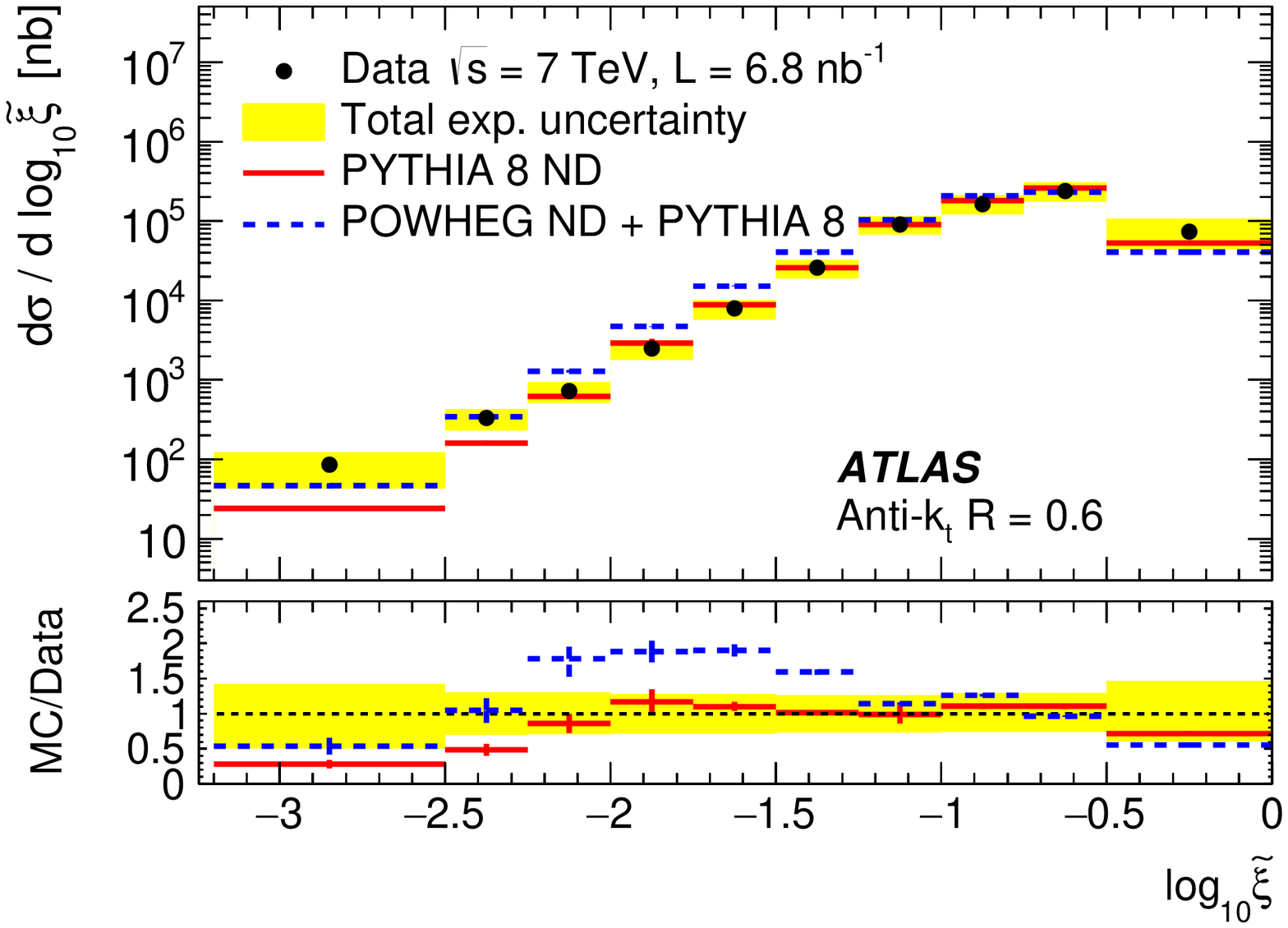}
        \caption{The dijet cross sections differential in (a) $\detaf$ and (b) 
$\xiepz$, compared with the PYTHIA8 ND MC model as well as an ND model using 
the NLO POWHEG generator with hadronisation based on PYTHIA8. Each of the 
models is separately normalised to match the data in the first $\detaf$ bin. 
The error bars on the data and the MC models indicate their respective 
statistical uncertainties, while the yellow bands show the total uncertainties
on the data. The lower panels show ratios of the MC models to the data where 
the error bars indicate the sum in quadrature of the statistical uncertainties 
arising from the data and the MC simulation.}
  \label{fig:NDcomp}
\end{center}
\end{figure}

The predicted ND contribution at large gap sizes is sensitive to the modelling 
of rapidity and transverse momentum fluctuations in the hadronisation
process, which are not yet well constrained. To establish the
presence of a diffractive contribution, it is therefore necessary to 
investigate the likely range of ND predictions. In Fig.~\ref{fig:NDcomp}, the 
dijet cross sections differential in \detaf and $\xiepz$ are compared
with the PYTHIA8 ND contribution and also with an NLO calculation of
non-diffractive dijet production in the POWHEG framework, with 
hadronisation modelled using PYTHIA8, as
described in Section~\ref{sec:MC}. Each of
the ND predictions is separately
normalised in the first bin of the $\detaf$
distribution. The range spanned by the 
ND predictions suggests a substantial uncertainty
in the probability of producing gaps through hadronisation fluctuations, 
such that for $\detaf \lesssim 4$,
it is not possible to draw conclusions on the presence or absence of an 
additional diffractive contribution. However, in both of the models, the
ND prediction falls significantly short of the data 
for $\detaf \gtrsim 4$. A similar conclusion is reached at the lowest $\xiepz$.
This region is therefore investigated in more detail in the following. 

\begin{figure}[h!]
  \begin{center}
\includegraphics[width=0.49\textwidth]{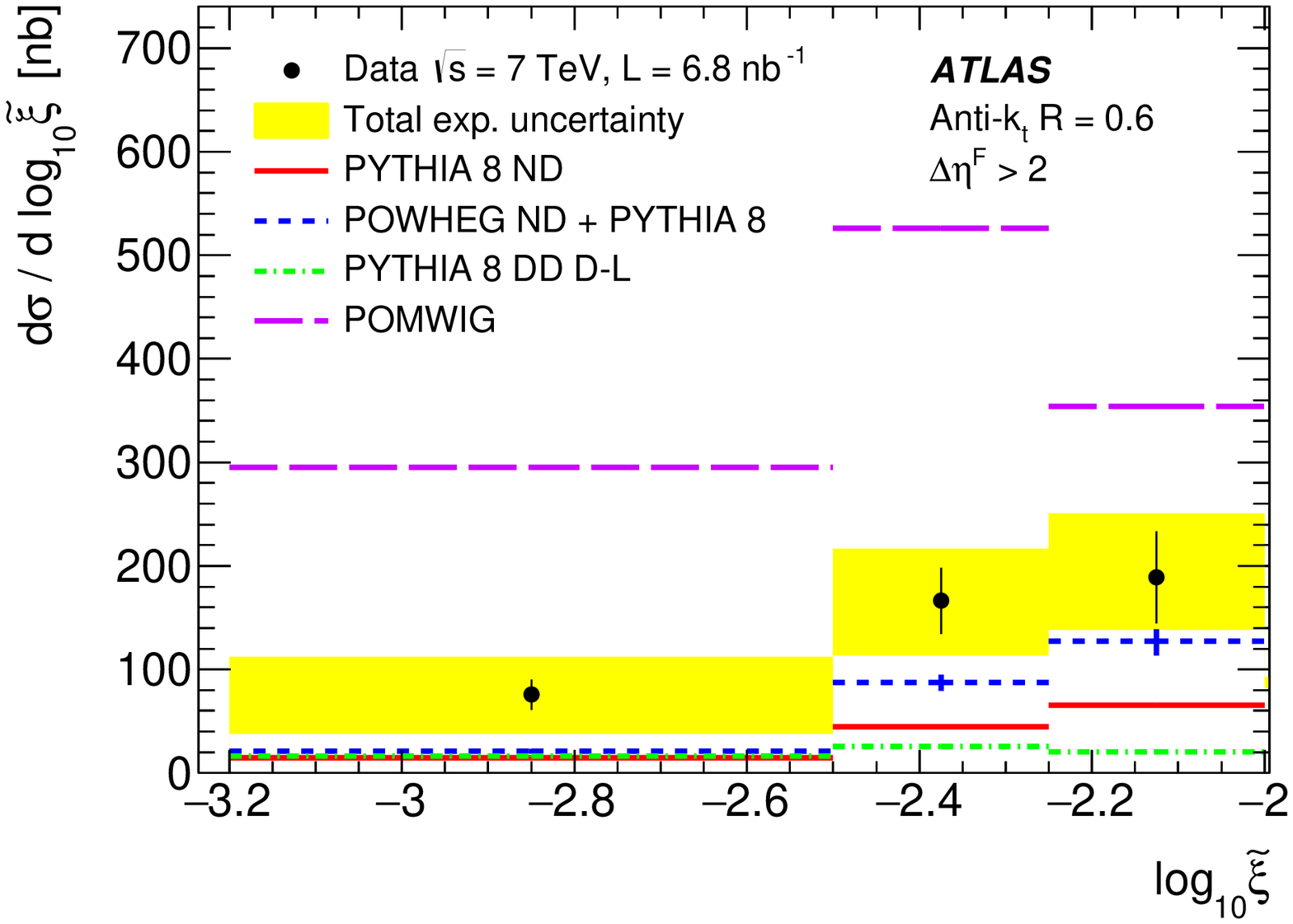}      
\includegraphics[width=0.49\textwidth]{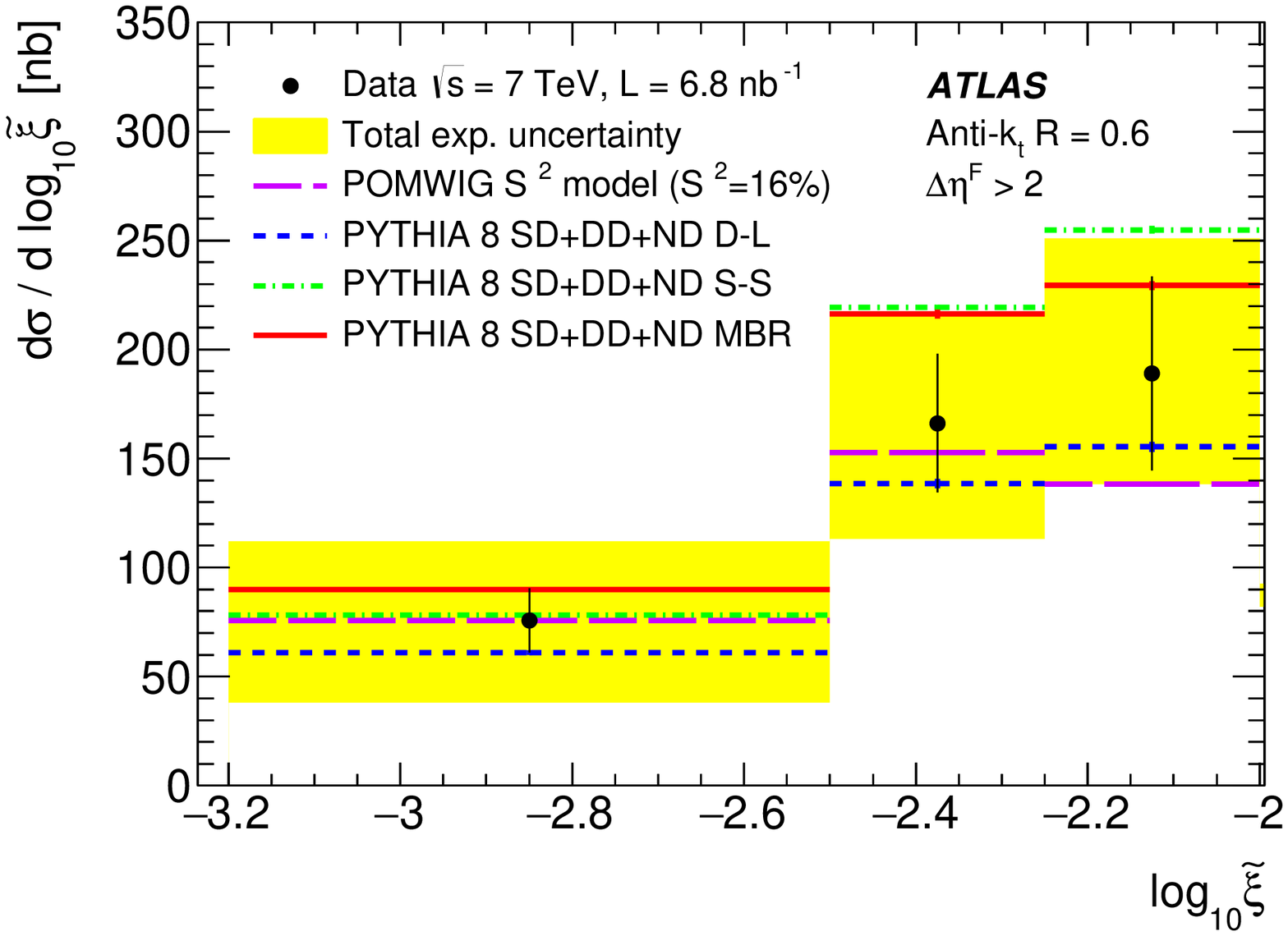}
  \caption{ The differential cross section as a function of $\xiepz$ 
for events satisfying $\detaf > 2$. The same data are shown in (a) and (b),
and are compared with models as described in the text. The error bars on the 
data and the MC models indicate their respective statistical uncertainties, 
while the yellow bands show the total uncertainties on the data.
The 'POMWIG $S^2$' model represents the sum of PYTHIA ND and POMWIG, with POMWIG
multiplied by $0.16$ and scaled by $1/1.23$ and by the (SD+DD)/SD ratio from 
PYTHIA8.} 
  \label{fig:results_Total_xiGaps3}
\end{center}
\end{figure}

Since the diffractive contribution is characterised by both large
$\detaf$ and small $\xiepz$, it can be separated most cleanly by 
placing requirements on 
both variables simultaneously. In Fig.~\ref{fig:results_Total_xiGaps3},
the $\xiepz$ distribution is shown after applying the requirement
$\detaf > 2$. This restricts the accessible kinematic range to
$\xiepz \lesssim 0.01$, and
suppresses the ND
contributions considerably. As shown in 
Fig.~\ref{fig:results_Total_xiGaps3}(a),
the ND contribution in the lowest
$\xiepz$ bin ($-3.2 < \log_{10} \xiepz < -2.5$)
is smaller than 25\%
according to all models considered, 
allowing for a quantitative investigation
of the diffractive contribution.

The data are compared with various models of diffractive dijet production with 
no rapidity-gap survival probability factors applied. The PYTHIA8 ND+SD+DD model
is shown in Fig.~\ref{fig:results_Total_xiGaps3}(b) for three different 
choices of pomeron flux, Schuler--Sj\"ostrand (S-S), Donnachie-Landshoff (D-L) 
and Minimum Bias Rockefeller (MBR), as described in Section~\ref{sec:MC}. 
The SD contribution dominates in this kinematic region, as can be inferred by 
comparing the PYTHIA8 predictions in Fig.~\ref{fig:results_Total_xiGaps3}(b)
with the PYTHIA8 ND and PYTHIA8 DD contributions in 
Fig.~\ref{fig:results_Total_xiGaps3}(a). There is some dependence of the 
predicted cross section on the choice of flux, but all three PYTHIA8 
predictions are compatible with the data without the need for a rapidity-gap 
survival probability factor, the D-L flux giving the best description. In contrast, 
the POMWIG model of the SD contribution alone lies above the data by around
a factor of three in the low $\xiepz$, large $\detaf$ region
(Fig.~\ref{fig:results_Total_xiGaps3}(a)).

Both PYTHIA8 and POMWIG are based on implementations of DPDFs as measured at 
HERA. POMWIG is a straightforward implementation of a standard factorisable 
Pomeron model with standard matrix elements, specifically intended for use in 
comparison with diffractive hard scattering processes such as that measured in 
this paper.
PYTHIA8 is intended to describe diffraction inclusively. It contains a complex 
transition between the hard (DPDF-based) and soft models, and the corresponding 
mechanisms for generating final-state particles.
The large difference here between the predictions of
PYTHIA8 and POMWIG
may be a consequence of this difference in basic approach.
The quality of the description of the data by PYTHIA8 is not
altered significantly if the modelling of multi-particle interactions,
colour reconnections, or initial- or final-state radiation are varied. 

Attributing the POMWIG model's excess over the data in the most sensitive 
region to absorptive effects, the data are compared quantitatively with 
POMWIG to determine the rapidity-gap survival probability $S^2$ appropriate to 
this model. The value of $S^2$ is determined from the region where the poorly 
known ND contribution is smallest, i.e. integrated over the range 
$-3.2 < \log_{10} \xiepz < -2.5$ 
after imposing the rapidity-gap requirement $\detaf > 2$ as in 
Fig.~\ref{fig:results_Total_xiGaps3}. 
The estimate of $S^2$ is obtained from the ratio of data to 
the SD contribution in the POMWIG model after subtracting from the
data the ND contribution as modelled by PYTHIA8 and the DD 
contribution assuming the SD/(SD+DD) ratio from PYTHIA8. No gap survival factors
are applied to the subtracted ND and DD contributions. The size of
these corrections can be inferred from the PYTHIA8 ND and DD contributions
as indicated in Fig.~\ref{fig:results_Total_xiGaps3}(a). A correction factor 
$1.23 \pm 0.16$ \cite{Aktas:2006hx} is applied to $S^2$ to account for the fact 
that the H1 2006 Fit B DPDFs used in POMWIG include proton dissociation
contributions $ep \rightarrow eXY$ where the proton excitation has a mass
$M_{\mathrm Y} < 1.6$ \GeV, in addition to the SD process. 

The resulting extracted value of the rapidity-gap survival probability 
appropriate to the mixed POMWIG / PYTHIA8 model is
\begin{eqnarray*}
S^2 = 0.16 \pm 0.04 \ {\rm (stat.)} \pm 0.08 \ {\rm (exp.\ syst.)} \ ,
\end{eqnarray*}
where the statistical (stat.) and experimental systematic (exp. syst.)
uncertainties are propagated from the data. 
This model is shown as `POMWIG $S^2$ Model' in 
Fig.~\ref{fig:results_Total_xiGaps3}(b). 
No attempt has been made to fully assess the model-dependence uncertainty,
although changing the ND contribution in the extraction from PYTHIA8 to
POWHEG + PYTHIA8 results in an $S^2$ of 0.15
and indications from elsewhere \cite{Andreev:2015cwa, CMS-FWD-10-004} suggest
that $S^2$ might be smaller if NLO models were used. 
The result is compatible with the values of $0.12 \pm 0.05$ 
and $0.08 \pm 0.04$, obtained by CMS in LO and NLO analyses, respectively, 
using the region $0.0003 < \widetilde{\xi} < 0.002$
and a jet $R$ parameter of 0.5 \cite{CMS-FWD-10-004}. 
The result is also compatible with that obtained at
lower centre-of-mass energy at the Tevatron \cite{Affolder:2000vb}, 
which was re-evaluated in a subsequent NLO 
analysis \cite{Klasen:2009bi}
to be between 0.05 and 0.3, depending on 
the fraction of the pomeron momentum 
carried by the parton entering the hard scattering. 
Theoretical predictions for $S^2$ at the 
LHC \cite{Kaidalov:2003gy,Khoze:2000wk}  
are also compatible with the result here, 
although the predicted decrease with increasing centre-of-mass energy 
is not yet established.

\FloatBarrier

%%%%%%%%%%%%%%%%%%%%%%%%%%%%%%%%%%%%
%%%%%%%%%%%%%%%%%%%%%%%%%%%%%%%%%%%%
%   _____                 _           _                 
%  / ____|               | |         (_)                
% | |     ___  _ __   ___| |_   _ ___ _  ___  _ __  ___ 
% | |    / _ \| '_ \ / __| | | | / __| |/ _ \| '_ \/ __|
% | |___| (_) | | | | (__| | |_| \__ \ | (_) | | | \__ \
%  \_____\___/|_| |_|\___|_|\__,_|___/_|\___/|_| |_|___/
%                                                       
%%%%%%%%%%%%%%%%%%%%%%%%%%%%%%%%%%%%

\section{Conclusions}
\label{ch:conclusions}
An ATLAS measurement of the cross section for dijet production in association 
with forward rapidity gaps is reported, based on $6.8 \ {\rm nb^{-1}}$ low 
pile-up 7 \TeV\ pp collision data taken at the LHC in 2010. The data are 
characterised according to the size of the forward rapidity gap, 
quantified by \detaf and $\xiepz$, which for the single-diffractive case 
approximates the fractional longitudinal momentum loss of the scattered 
proton using the information available within the detector acceptance. 
Non-diffractive Monte Carlo models are capable of describing
the data over a wide kinematic range. 
However, a diffractive component is also required for a more complete 
description of the data, particularly when both large \detaf and small $\xiepz$ 
are required. The PYTHIA8 model gives the best description of the shape and 
normalisation of this contribution.

The rapidity-gap survival probability is estimated by comparing the measured 
cross section for events with both large $\detaf$ and small $\xiepz$ with the 
leading-order POMWIG Monte Carlo model of the diffractive contribution,
derived from diffractive parton distribution functions extracted in deep 
inelastic $ep$ scattering. This determination is limited by the uncertainties 
associated with the non-diffractive and double-dissociation
contributions, the result being 
$S^2 = 0.16 \pm 0.04 \ {\rm (stat.)} \pm 0.08 \ {\rm (exp. \ syst.)}$.

%-------------------------------------------------------------------------------
\section*{Acknowledgements}
%-------------------------------------------------------------------------------
\label{sec:acknowldge}
% Acknowledgements for papers with collision data
% Version 15-Sep-2015

% Standard acknowledgements start here
%----------------------------------------------
We thank CERN for the very successful operation of the LHC, as well as the
support staff from our institutions without whom ATLAS could not be
operated efficiently.

We acknowledge the support of ANPCyT, Argentina; YerPhI, Armenia; ARC, Australia; BMWFW and FWF, Austria; ANAS, Azerbaijan; SSTC, Belarus; CNPq and FAPESP, Brazil; NSERC, NRC and CFI, Canada; CERN; CONICYT, Chile; CAS, MOST and NSFC, China; COLCIENCIAS, Colombia; MSMT CR, MPO CR and VSC CR, Czech Republic; DNRF, DNSRC and Lundbeck Foundation, Denmark; IN2P3-CNRS, CEA-DSM/IRFU, France; GNSF, Georgia; BMBF, HGF, and MPG, Germany; GSRT, Greece; RGC, Hong Kong SAR, China; ISF, I-CORE and Benoziyo Center, Israel; INFN, Italy; MEXT and JSPS, Japan; CNRST, Morocco; FOM and NWO, Netherlands; RCN, Norway; MNiSW and NCN, Poland; FCT, Portugal; MNE/IFA, Romania; MES of Russia and NRC KI, Russian Federation; JINR; MESTD, Serbia; MSSR, Slovakia; ARRS and MIZ\v{S}, Slovenia; DST/NRF, South Africa; MINECO, Spain; SRC and Wallenberg Foundation, Sweden; SERI, SNSF and Cantons of Bern and Geneva, Switzerland; MOST, Taiwan; TAEK, Turkey; STFC, United Kingdom; DOE and NSF, United States of America. In addition, individual groups and members have received support from BCKDF, the Canada Council, CANARIE, CRC, Compute Canada, FQRNT, and the Ontario Innovation Trust, Canada; EPLANET, ERC, FP7, Horizon 2020 and Marie Skłodowska-Curie Actions, European Union; Investissements d'Avenir Labex and Idex, ANR, Region Auvergne and Fondation Partager le Savoir, France; DFG and AvH Foundation, Germany; Herakleitos, Thales and Aristeia programmes co-financed by EU-ESF and the Greek NSRF; BSF, GIF and Minerva, Israel; BRF, Norway; the Royal Society and Leverhulme Trust, United Kingdom.

The crucial computing support from all WLCG partners is acknowledged
gratefully, in particular from CERN and the ATLAS Tier-1 facilities at
TRIUMF (Canada), NDGF (Denmark, Norway, Sweden), CC-IN2P3 (France),
KIT/GridKA (Germany), INFN-CNAF (Italy), NL-T1 (Netherlands), PIC (Spain),
ASGC (Taiwan), RAL (UK) and BNL (USA) and in the Tier-2 facilities
worldwide.
%----------------------------------------------

%%%%%%%%%%%%%%%%%%%%%%%%%%%%%%%%%%%%
%  ____  _ _     _ _                             _           
% |  _ \(_) |   | (_)                           | |          
% | |_) |_| |__ | |_  ___   __ _ _ __ __ _ _ __ | |__  _   _ 
% |  _ <| | '_ \| | |/ _ \ / _` | '__/ _` | '_ \| '_ \| | | |
% | |_) | | |_) | | | (_) | (_| | | | (_| | |_) | | | | |_| |
% |____/|_|_.__/|_|_|\___/ \__, |_|  \__,_| .__/|_| |_|\__, |
%                           __/ |         | |           __/ |
%                          |___/          |_|          |___/ 
%
%%%%%%%%%%%%%%%%%%%%%%%%%%%%%%%%%%%%
\printbibliography

\newpage
% ATLAS Collaboration author list
% Data extracted on 22-Aug-2015 for paper reference STDM-2014-04
%\documentclass[11pt]{article}
%\usepackage{a4wide}\begin{document}
\begin{flushleft}
{\Large The ATLAS Collaboration}

\bigskip

G.~Aad$^{\rm 85}$,
B.~Abbott$^{\rm 113}$,
J.~Abdallah$^{\rm 151}$,
O.~Abdinov$^{\rm 11}$,
R.~Aben$^{\rm 107}$,
M.~Abolins$^{\rm 90}$,
O.S.~AbouZeid$^{\rm 158}$,
H.~Abramowicz$^{\rm 153}$,
H.~Abreu$^{\rm 152}$,
R.~Abreu$^{\rm 116}$
% ...
\bigskip
\\
% ...
$^{11}$ Institute of Physics, Azerbaijan Academy of Sciences, Baku, Azerbaijan\\
% ...
$^{85}$ CPPM, Aix-Marseille Universit{\'e} and CNRS/IN2P3, Marseille, France\\
% ...
$^{90}$ Department of Physics and Astronomy, Michigan State University, East Lansing MI, United States of America\\
% ...
$^{107}$ Nikhef National Institute for Subatomic Physics and University of Amsterdam, Amsterdam, Netherlands\\
% ...
$^{113}$ Homer L. Dodge Department of Physics and Astronomy, University of Oklahoma, Norman OK, United States of America\\
% ...
$^{116}$ Center for High Energy Physics, University of Oregon, Eugene OR, United States of America\\
% ...
$^{151}$ Institute of Physics, Academia Sinica, Taipei, Taiwan\\
$^{152}$ Department of Physics, Technion: Israel Institute of Technology, Haifa, Israel\\
$^{153}$ Raymond and Beverly Sackler School of Physics and Astronomy, Tel Aviv University, Tel Aviv, Israel\\
% ...
$^{158}$ Department of Physics, University of Toronto, Toronto ON, Canada\\
% ...
\end{flushleft}

%\end{document}
% Created with xml2latex.py

%%%%%%%%%%%%%%%%%%%%%%%%%%%%%%%%%%%%
%  ______           _   _____                                        _   
% |  ____|         | | |  __ \                                      | |  
% | |__   _ __   __| | | |  | | ___   ___ _   _ _ __ ___   ___ _ __ | |_ 
% |  __| | '_ \ / _` | | |  | |/ _ \ / __| | | | '_ ` _ \ / _ \ '_ \| __|
% | |____| | | | (_| | | |__| | (_) | (__| |_| | | | | | |  __/ | | | |_ 
% |______|_| |_|\__,_| |_____/ \___/ \___|\__,_|_| |_| |_|\___|_| |_|\__|
%                                                                        
%%%%%%%%%%%%%%%%%%%%%%%%%%%%%%%%%%%%

\end{document}